\documentclass{article}
\usepackage{times}
\usepackage{amsfonts}
\usepackage{graphicx}
\usepackage[pdfmark]{hyperref}
\begin{document}
\noindent
{\Large POSITIVE CURVATURE CAN MIMIC A QUANTUM}
\vskip1cm
\noindent
{\bf J.M. Isidro}${}^{1,a}$, {\bf J.L.G. Santander}${}^{2,b}$ and {\bf P. Fern\'andez de C\'ordoba}${}^{1,c}$ \\
${}^{1}$Instituto Universitario de Matem\'atica Pura y Aplicada,\\ Universidad Polit\'ecnica de Valencia, Valencia 46022, Spain\\
${}^{2}$Departamento de Ciencias Experimentales y Matem\'aticas,\\ Universidad Cat\'olica de Valencia, Valencia 46002, Spain\\
${}^{a}${\tt joissan@mat.upv.es}, ${}^{b}${\tt jlgonzalez@mat.upv.es}, \\
${}^{c}${\tt pfernandez@mat.upv.es}

\vskip1cm
\noindent
{\bf Abstract} We elaborate on the existing idea that quantum mechanics is an emergent phenomenon, in the form of a coarse--grained description of some underlying deterministic theory. We apply the Ricci flow as a technical tool to implement dissipation, or information loss, in the passage from an underlying deterministic theory to its emergent quantum counterpart. A key ingedient in this construction is the fact that the space of physically inequivalent quantum states (either pure or mixed) has positive Ricci curvature. This leads us to an interesting thermodynamical analogy of emergent quantum mechanics.

\section{Introduction}\label{intt}

Quantum mechanics as a statistical theory has been argued to {\it emerge}\/ from an underlying deterministic theory \cite{THOOFT2}. Specifically, for any quantum system there exists at least one deterministic model that reproduces all its dynamics after prequantisation. This existence theorem has been extended  to include cases characterised by sets of commuting beables \cite{ELZE1}; it  has also been complemented with an explicit dynamical theory \cite{ELZE5}. 

Mechanisms have been presented \cite{THOOFT2, ELZE1, ELZE5} to explain the passage from a deterministic theory to a probabilistic theory. Usually they are based on a dynamical system, the phase--space trajectories of which possess suitably located attractors ({\it e.g.}, at the eigenvalues of the given quantum Hamiltonian, or at certain configurations of the density matrix).  These mechanisms can be thought of as an existence theorem, in that every quantum system (with a finite--dimensional Hilbert space) possesses at least one deterministic system underlying it. 

On the other hand there are plenty of dissipation equations in physics and mathematics, equations implementing the information loss that is characteristic of the passage from classical to quantum. The heat equation immediately comes to mind. 

In this contribution we develop a deterministic model exhibiting dissipation, from which quantum mechanics emerges naturally. Given a quantum mechanics with a complex $d$--dimensional Hilbert space, the Lie group $SU(d)$ represents classical canonical transformations on the projective space $\mathbb{CP}^{d-1}$ of quantum states.  Let $R$ stand for the Ricci flow \cite{TOPPING} of the manifold $SU(d-1)$ down to one point, and let $P$ denote the projection from the Hopf bundle onto its base $\mathbb{CP}^{d-1}$. Then the underlying deterministic model we propose here is the Lie group $SU(d)$, acted on by the operation $PR$. 

We would like to mention that additional quantum--mechanical applications of the Ricci flow have been reported in \cite{CARROLL1, CARROLL2, ISIDRO1, ISIDRO2}; deterministic models of quantum mechanics and closely related topics are dealt with at length in \cite{DICE, ANDREI, MATONE1, MATONE2}.

\section{The Ricci flow as a (nonlinear) heat flow}\label{grund}

Given an $n$--dimensional manifold $\mathbb{M}$ endowed with the Riemannian metric $g_{ij}$, the equation governing the (unnormalised) Ricci flow reads
\begin{equation}
\frac{\partial g_{ij}}{\partial t}=-2R_{ij}, \qquad i,j=1,\ldots, n,\qquad t\geq 0,
\label{guillaume}
\end{equation}
where $t$ is an evolution parameter (not a coordinate on $\mathbb{M}$), and $R_{ij}$ is the Ricci tensor corresponding to the metric $g_{ij}$. Informally one can say that Ricci--flat spaces remain unchanged under the flow, while positively curved manifolds contract  and negatively curved manifolds expand under the flow. We will be interested in the particular case of Einstein manifolds, where the Ricci tensor and the metric are proportional:
\begin{equation}
R_{ij}=\kappa g_{ij}, 
\label{onestone}
\end{equation}
with $\kappa$ a constant.  Since the metric $g_{ij}$ is assumed positive definite,  the sign of $\kappa$ equals the sign of the Ricci tensor. Relevant examples of positively curved Einstein manifolds are complex projective space $\mathbb{CP}^N$ and the special unitary group $SU(N)$, both of which will play an important role in what follows. Their respective metrics are the Fubini--Study metric \cite{KOBAYASHI} and the Killing--Cartan metric \cite{HELGASON}. 

Under the Ricci flow, the contraction of a whole manifold down to a point can play the role of a dissipative mechanism. One hint that this intuition is correct comes from the following example. Consider a 2--dimensional manifold endowed with the isothermal coordinates $x$ and $y$. Then the metric reads 
\begin{equation}
{\rm d}s^2={\rm e}^{-f(x,y)}\left({\rm d}x^2+{\rm d}y^2\right).
\label{isso}
\end{equation}
Allowing the metric to depend also on the evolution parameter $t$, the Ricci flow equation (\ref{guillaume}) becomes
\begin{equation}
\frac{\partial f}{\partial t}=\nabla^2 f.
\label{heiss}
\end{equation}
The above is formally identical to the heat equation, with one important difference, however: the Laplacian $\nabla^2$ is computed with respect to the metric (\ref{isso}), in which it reads
\begin{equation}
\nabla^2 f={\rm e}^{f}\left(\frac{\partial^2 f}{\partial x^2}+\frac{\partial^2f}{\partial y^2}\right).
\label{placa}
\end{equation}
Regardless of the nonlinearity of (\ref{placa}), the fact that the Ricci--flow equation can be recast as a generalisation of the heat equation is a clear hint that a dissipative mechanism is at work.

\section{The deterministic model for pure states}\label{fluss}

In this section we will consider a quantum system with a finite, complex $d$--dimensional Hilbert space of quantum states, that we can identify with $\mathbb{C}^d$. Let ${\cal C}$ denote the phase space of the classical model, the quantisation of which gives the quantum system under consideration. For our purposes the precise nature of this classical model on ${\cal C}$ is immaterial. Now unitary transformations on Hilbert space are the quantum counterpart of canonical transformations on classical phase space ${\cal C}$. We may thus regard $SU(d)$ as representing classical canonical transformations,  $\mathbb{C}^d$ being the carrier space of this representation. We are considering, as  in ref. \cite{THOOFT2}, the simplified case of a finite--dimensional Hilbert space. Without loss of generality we will restrict to those canonical transformations that are represented by unitary matrices with determinant equal to 1.

Now quantum states are unit rays rather than vectors, so in fact the true space of inequivalent quantum states is the complex projective space $\mathbb{CP}^{d-1}$.  The latter can be regarded as a homogeneous manifold: 
\begin{equation}
\mathbb{CP}^{d-1}=\frac{SU(d)}{SU(d-1)\times U(1)}.
\label{homogen}
\end{equation}
In this picture we have $SU(d)$ as the total space of a fibre bundle with typical fibre $SU(d-1)\times U(1)$ over the base manifold $\mathbb{CP}^{d-1}$. The projection map 
\begin{equation}
\pi:SU(d)\longrightarrow\mathbb{CP}^{d-1}, \qquad \pi(w):=[w]
\label{projektion}
\end{equation}
arranges points $w\in SU(d)$ into $SU(d-1)\times U(1)$ equivalence classes $[w]$. 

Classical canonical transformations as represented by $SU(d)$ act on the Hilbert space $\mathbb{C}^d$. This descends to an action $\alpha$ of $SU(d)$ 
on $\mathbb{CP}^{d-1}$ as follows:
\begin{equation}
\alpha:SU(d)\times\mathbb{CP}^{d-1}\longrightarrow\mathbb{CP}^{d-1},\quad \alpha\left(u,[v]\right):=[uv].
\label{yeste}
\end{equation}
Here we have $u\in SU(d)$, $[v]\in \mathbb{CP}^{d-1}$, and $uv$ denotes $d\times d$ matrix multiplication.
One readily checks that this action is well defined on the equivalence classes under right multiplication by elements of the stabiliser subgroup $SU(d-1)\times U(1)$. 
This allows one to regard quantum states as equivalence classes of classical canonical transformations on ${\cal C}$. Physically, $u$ in (\ref{yeste}) denotes (the representative matrix of) a canonical transformation on ${\cal C}$, and $[v]$ denotes the equivalence class of (representative matrices of) the canonical transformation $v$ or, equivalently, the quantum state $\vert v\rangle$.  

In the picture just sketched, two canonical transformations are equivalent whenever they differ by a canonical transformation belonging to $SU(d-1)$, and/or whenever they differ by a $U(1)$--transformation. Modding out by $U(1)$ has a clear physical meaning: it is the standard freedom in the choice of the phase of the wavefunction corresponding to the matrix $v\in SU(d)$. Modding out by $SU(d-1)$ also has a physical meaning: canonical transformations on the $(d-1)$--dimensional subspace  $\mathbb{C}^{d-1}\subset\mathbb{C}^d$ are a symmetry of $v$. Therefore the true quantum state $\vert v\rangle$ is obtained from $v\in SU(d)$ after modding out by the stabiliser subgroup $SU(d-1)\times U(1)$.

We conclude that this picture contains some of the elements identified as responsible for the passage from a classical world (canonical transformations) to a quantum world (equivalence classes of canonical transformations, or unit rays within Hilbert space). This is so because some kind of  dissipative mechanism is at work, through the emergence of orbits, or equivalence classes. However the projection (\ref{projektion}) is an on/off mechanism. Instead, one would like to see dissipation occurring as a flow along some continuous parameter. To this end we need some deterministic flow governed by some differential equation. 

We claim that we can render the projection (\ref{projektion}) a dissipative mechanism governed by some differential equation. This equation will turn out to be the Ricci flow (\ref{guillaume}). Proof of this statement follows.

The Lie group $SU(d-1)\times U(1)$ is compact, but it is not semisimple due to the Abelian factor $U(1)$. Leaving the $U(1)$ factor momentarily aside, $SU(d-1)$ is semisimple and compact. As such it qualifies as an Einstein space with positive scalar curvature with respect to the Killing--Cartan metric \cite{HELGASON}. Now eqn. (\ref{guillaume}) ensures that $SU(d-1)$ contracts to a point under the Ricci flow.

However  the $U(1)$ factor renders $SU(d-1)\times U(1)$ nonsemisimple. As a consequence, the Killing--Cartan metric of $SU(d-1)\times U(1)$ has a vanishing determinant \cite{HELGASON}. The Ricci flow can still cancel the $SU(d-1)$--factor within $SU(d)$, but not the $U(1)$ factor. After contracting $SU(d-1)$ to a point we are left with the space $U(1)\times \mathbb{CP}^{d-1}$ or, more generally, with a $U(1)$--bundle over the base manifold $\mathbb{CP}^{d-1}$. This $U(1)$--bundle over $\mathbb{CP}^{d-1}$ is the Hopf bundle, where the total space is the sphere $S^{2d-1}$ in $2d-1$ real dimensions \cite{KOBAYASHI}. This sphere falls short of being the true space of quantum states by the unwanted $U(1)$--fibre, that cannot be removed by the Ricci flow. It can, however, be done away with by projection $P$ from the total space of the bundle down to its base. The combined operation ``Ricci flow $R$, followed by projection $P$" acts on the stabiliser subgroup $SU(d-1)\times U(1)$ of the initial $SU(d)$ and leaves us with $\mathbb{CP}^{d-1}$ as desired. Therefore this combined operation $PR$ acts in the same way as the projection $\pi$ in (\ref{projektion}). As opposed to the latter, however, this combined operation $PR$ provides us with a  differential equation that implements dissipation along a continuous parameter, at least along most of the way.

\section{The deterministic model for mixed states}\label{misch}

We have so far dealt only with pure quantum states.  In trying to extend our previous analysis to mixed quantum states, we must first answer the following two questions: What is the manifold of mixed quantum states, and what sign does its Ricci scalar have? Mixed states can be represented by density matrices $D$, expressible as 
\begin{equation}
D=\sum_{j=1}^n\vert v_j\rangle p_j\langle v_j\vert,\qquad n>1,
\label{ketefollenramallokabron}
\end{equation}
the $p_j> 0$ being the probability of finding the system in the pure state $\vert v_j\rangle$. Above we assume that $n>1$, {\it i.e.}, that the state considered is not pure but truly mixed. The $p_j$ must add up to unity,
\begin{equation}
\sum_{j=1}^np_j=1.
\label{traza}
\end{equation}
It turns out that the manifold of density matrices is a norm--closed (with respect to the trace norm), convex subset of the unit sphere of the space of trace--class operators  (see, {\it e.g.}, \cite{THIRRING}). In the finite--dimensional setup considered here, all operators are trace class, and we are left with a convex subset of the unit sphere of the space of $d\times d$ Hermitian matrices. Now the space of $d\times d$ Hermitian matrices has (real) dimension $d^2-1$, so its unit sphere is $d^2-2$ (real) dimensional. The manifold of mixed states is a convex subset of the real sphere $S^{d^2-2}$.  In particular, the latter has positive Ricci curvature. On the other hand, the $N$--dimensional sphere $S^N$ equals the homogeneous manifold $SO(N+1)/SO(N)$, so our manifold of mixed quantum states is a convex subset of
\begin{equation}
S^{d^2-2}=\frac{SO(d^2-1)}{SO(d^2-2)}.
\label{moho}
\end{equation}
As in the case of pure states, the special orthogonal groups in the numerator and in the denominator above carry positive Ricci curvature with respect to the corresponding Killing--Cartan metric \cite{HELGASON}. 

Now eqn. (\ref{moho}) differs very little from (\ref{homogen}), that we discussed at length in the case of pure quantum states. One difference between these two equations is that (\ref{homogen}) contains the unitary groups, while (\ref{moho}) contains the special orthogonal groups. Another difference is that the Abelian factor $U(1)$ in the denominator of (\ref{homogen}) has disappeared from (\ref{moho}) (one could still mod out the right--hand side of (\ref{moho}) by the discrete group $\mathbb{Z}_2$ in order to obtain real projective space, but the latter is not related to the space of density matrices). One final difference between the pure and the mixed case is that the latter does not have the full left--hand side of (\ref{moho}) as the space of quantum states, but only a convex subset thereof.

All this  notwithstanding, these three differences do not suffice to prevent the analysis (and the ensuing conclusions) of the case of pure quantum states from applying to the case of mixed quantum states as well. It is interesting to observe that mixed quantum states are actually simpler to deal with than pure states, because the absence of the Abelian factor in the denominator of (\ref{moho}) allows one to dispense with the projection $P$ from the Hopf bundle---in fact, in the mixed case there is no Hopf bundle at all.

Having seen that the case of mixed quantum states does not differ substantially from that of pure states, for the rest of this contribution we will concentrate on pure quantum states.

\section{Positive curvature mimics a quantum}\label{unterhaltung}

Our starting point was the observation that canonical transformations on classical phase space are implemented quantum--mechanically as unitary transformations on the Hilbert space of quantum states. In our finite--dimensional setup,  this gave rise to a natural action of $SU(d)$ on $\mathbb{C}^d$. This action  provided us with the building blocks to construct the deterministic system that we take to underlie the given quantum mechanics. Next, different pieces of classical information (elements of $SU(d)$, or classical canonical transformations) were arranged into quantum equivalence classes (points on $\mathbb{CP}^d$, or quantum states): this procedure implements information loss, or dissipation. Quantum states thus arose as equivalence classes of canonical transformations on classical phase space. However, dissipation was not implemented by means of the usual projection (\ref{projektion}) (an on/off mechanism), but rather by means of the Ricci flow (followed by the projection $P$). The rationale was that the Ricci flow provided us with a a deterministic mechanism governed by a dissipative differential equation, that can be understood as a flow along a continuous parameter. 

In a nutshell, our deterministic model is the group manifold $SU(d)$, acted on by the combined operation $PR$ described above. Here $R$ stands for the Ricci flow of $SU(d-1)$ down to one point, and $P$ stands for the projection from the Hopf bundle with total space $S^{2d-1}$ onto its base $\mathbb{CP}^{d-1}$.  

The previous conclusions can be compactly recast, somewhat in the style of newspaper headlines, as {\it positive curvature mimics a quantum}\/ \cite{KOCH1, KOCH2}. This raises the question, how about negative curvature? From what was said above it should be clear that the answer is negative:  negative curvature {\it cannot}\/ mimic a quantum, since negative curvature causes expansion, rather than contraction to a point. Temporarily abandoning our geometrical stance, let us take a brief thermodynamical detour that will lend further support to our statement concerning negative curvature. Thermodynamics is a coarse--grained version of an underlying microscopic theory, namely, statistical mechanics. Coarse--graining the notion of mechanical energy in statistical mechanics gives rise to the thermodynamical notion of heat. In our setup, the analogue of the heat equation is the Ricci flow equation. Admittedly, the classical heat equation is linear while the Ricci flow equation is not, but this difference will play no role here. In fact, the analogy between these two equations goes so far, that the Ricci flow equation (\ref{guillaume}) can actually be derived from a functional ${\cal F}$, called Perelman's functional \cite{TOPPING}, which happens to be a monotonically increasing function of the time $t$. Therefore ${\cal F}$ qualifies as an entropy. Now heat flow occurs from higher temperature to lower temperature; such a heat flow is accompanied by an increase in entropy. This reveals what the thermodynamical analogue of negative curvature must be: heat flowing from {\it lower}\/ temperature to {\it higher}\/ temperature. This is clearly unphysical. Thus thermodynamics meets geometry in the statement that negative curvature cannot mimic a quantum. 

We would also like to point out that a related form of coarse--graining has been put forward \cite{ELZE6} in order to explain the emergence of quantum mechanics from an underlying deterministic theory. In fact we can provide a dictionary between our thermodynamical analogy, on the one hand,  and the requirements imposed on the operation of coarse--graining in ref. \cite{ELZE6}, on the other. Namely: probability conservation in \cite{ELZE6} corresponds to energy conservation in our thermodynamical analogy, while dissipation in \cite{ELZE6} is matched, in our picture, by increase in entropy.

Looking beyond, one could even pose the question: regardless of the sign, {\it why curvature at all?}\/ We have seen that Ricci--flat spaces remain unchanged under the Ricci flow. In our thermodynamical analogy, this would correspond to no heat flow at all, that is, to a temperature distribution satisfying the static Laplace equation $\nabla^2T=0$. A nonvanishing (and, as we have argued, positive) value of the scalar curvature provides us with a natural length scale: the Ricci scalar. With it, a natural notion of a quantum comes along.

\vskip2cm
\noindent
{\bf Acknowledgements} J.M.I. is pleased to thank the organizers of the meeting QTRF5 (Quantum Theory: Reconsideration of Foundations 5, V\"axj\"o, Sweden) for the invitation to participate, and for providing a congenial environment for scientific exchange. It is also a great pleasure to thank H.-T. Elze and B. Koch for discussions. This work has been supported by Universidad Polit\'ecnica de Valencia under grant PAID-06-09.---{\it Once you've reached the top, don't forget to send the elevator down, for the next guy.}

\end{document}